\documentclass[fleqn,twoside]{article}
\usepackage{espcrc2}
\usepackage{graphicx}

\newcommand{\AmS}{{\protect\the\textfont2
  A\kern-.1667em\lower.5ex\hbox{M}\kern-.125emS}}

  \def\rys#1#2{\begin{figure}[htb]
      \vskip 3mm
      \centerline{
      \includegraphics*[width=0.5 \textwidth]{#1}
      }
      \caption{#2}
      \vskip 3mm
      \end{figure}
      }

\title{Comparison of predictions for nuclear effects in the Marteau model with the NUX+FLUKA scheme}

\author{Jaros\l aw A. Nowak,\address[IFT] {Institute of Theoretical
Physics, Wroc\l aw
University.\\
pl. M. Borna 9, 50-204 Wroc\l aw, Poland}
\thanks{All the authors are supported by KBN
grant 105/E-344/SPB/ICARUS/P-03/DZ211/2003-2005} Cezary Juszczak
\addressmark[IFT], Jan~T.~Sobczyk\addressmark[IFT]}

\begin{document}

\begin{abstract}
Nuclear effects in neutrino-nucleus reactions simulated by means
of the NUX+FLUKA Monte Carlo generator are compared with the
theoretical predictions of the Marteau model. Pion absorption in
NUX+FLUKA and non-pionic $\Delta$ decays in the Marteau model
differ by about 30\%. The fraction of pions produced due to the
re-interactions after primary quasi-elastic vertex is in the
NUX+FLUKA scheme much higher then provided by the Marteau model.
\vspace{10mm}
\end{abstract}

\maketitle

\section{INTRODUCTION}

The aim of our investigation was to compare predictions for
neutrino interactions provided by NUX+FLUKA Monte Carlo scheme
\cite{nuxfluka} with those of the Marteau model \cite{marteau}. We
were interested mostly in nuclear effects in pion production but
for completeness (and for the sake of checking normalization
factors) we have also presented the results for
the quasi-elastic reaction.\\

The motivation for this study can be summarized in the following
points:
\medskip

i) In  recent years there has been growing interest  in the
studies of neutrino interactions at energies of a few GeV where
the dominant contributions to the cross section come from
quasi-elastic and single pion production channels \cite{nuint}.
\smallskip

ii) It is well known that the NUX+FLUKA MC scheme designed
originally to describe neutrino interactions at higher energies
is not satisfactory in a few GeV neutrino energy region
\cite{zeller}. It does not include separate resonance
contribution implemented in alternative MC codes after
Rein-Sehgal model \cite{reinsehgal}.
\smallskip

iii) Developments in quark-hadron duality suggest that perhaps it
is not necessary to include many resonances apart from the
$\Delta$ excitation \cite{bodek,casper}.
\smallskip

iv) The Marteau model describing together quasi-elastic and
$\Delta$ production reactions on nuclear targets is an interesting
candidate for implementation in MC codes. Its original version
includes a contribution from 2 particles - 2 holes excitations.
The model was a subject of further investigation \cite{sobczyk}
and it is important to understand if it is useful for practical
applications.
\smallskip

v) On the purely theoretical side there are interesting questions
concerning the way in which nuclear effects beyond Fermi gas model
are taken into account. One way is to describe them numerically
e.g. by FLUKA. An alternative is to perform sophisticated
theoretical computations like RPA. It is important to compare
predictions of these two approaches.\\

\rys{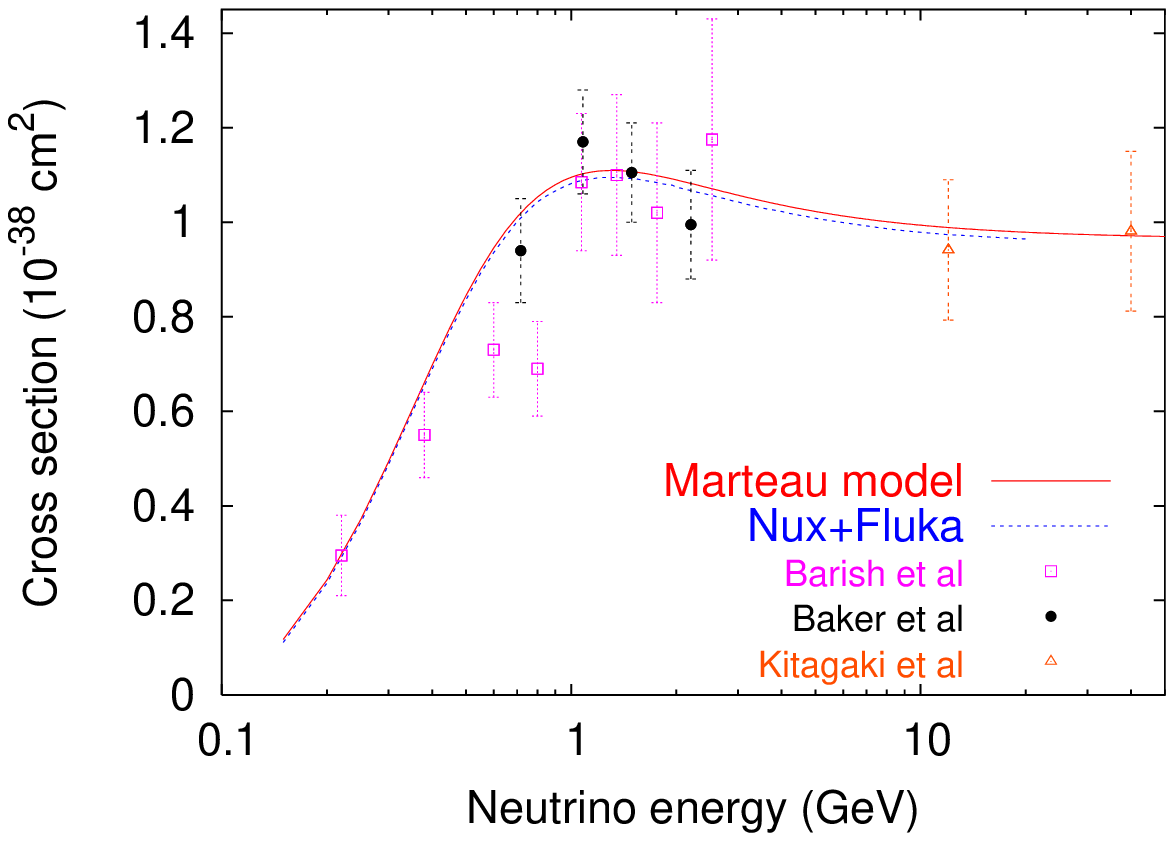}{The total cross section for the CC quasi-elastic
$\nu_{\mu}$ scattering on free nucleon. The experimental points
are taken from \cite{qel_exp}. \label{com_qel_free}}

\rys{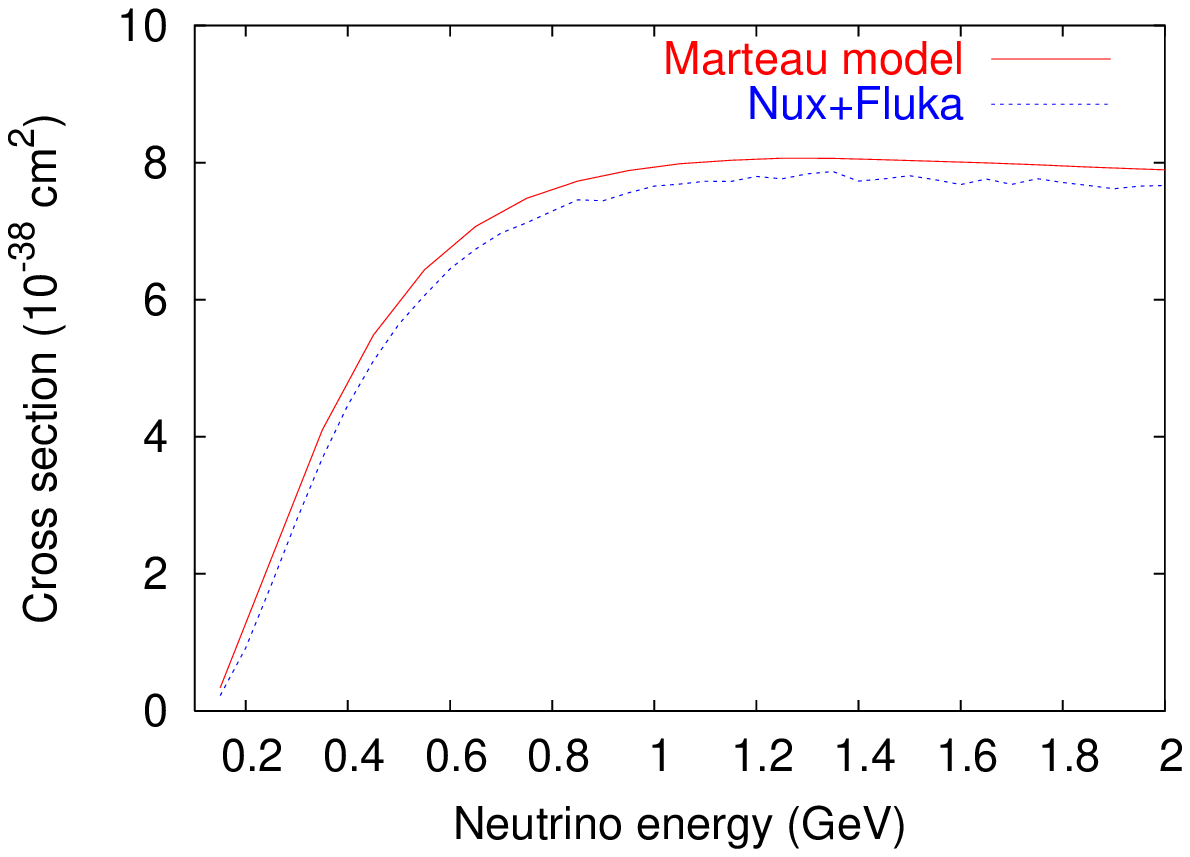}{The total cross section for the CC
quasi-elastic $\nu_{\mu}$ scattering on $^{16}O$ without Final
State Interactions.\label{com_qel_FG}}

\rys{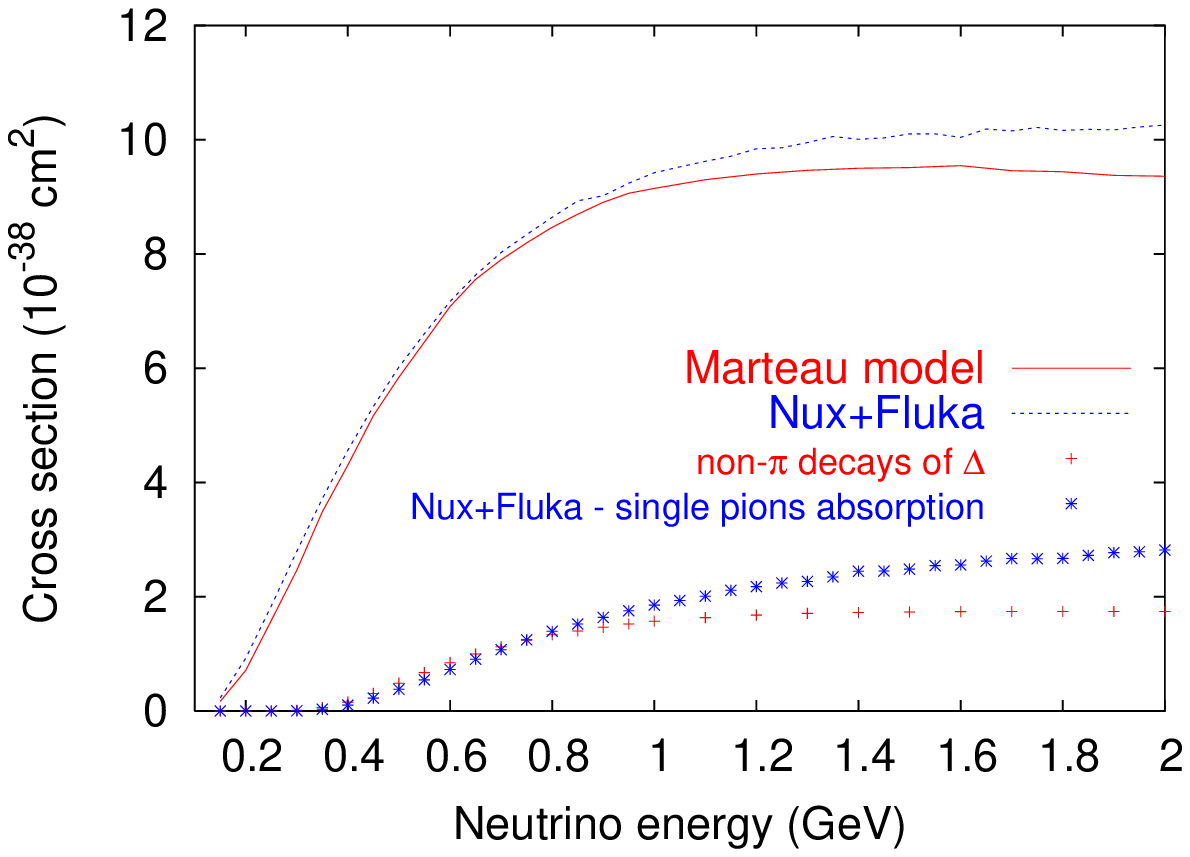}{The cross section for the CC $\nu_{\mu}$
quasi-elastic scattering  on $^{16}O$ (no pions in the final
state). The contributions from pion absorption (NUX+FLUKA) and
non-pion $\Delta$ resonance decays (Marteau model) are shown
separately with the dotted lines.\label{com_qel_RPA}}

In this paper we confine ourselves to the CC reactions of muon
neutrinos. The NUX+FLUKA Monte Carlo predictions were made by
using the PRET driver. For the total cross sections plots a
sample of $10^5$ events was produced for each neutrino energy
value. The events were classified based on some primary vertex
characteristics as well as the number of pions in the final
state. The overall normalization factor for the cross section was
taken from the PRET's output file pre1.out.

The differential cross section, i.e. the hadronic invariant mass
distribution in the $\nu p\rightarrow \mu^- p\pi^+$ channel, was
obtained from  a sample of $10^6$ inelastic events with the
neutrino energy fixed at 1GeV.\\

All the numerical predictions of the Marteau model are based on
\cite{sobczyk}.

\section{QUASI-ELASTIC REACTIONS}

The total cross sections for CC quasi-elastic scattering on free
nucleons are compared in Fig.~1. We find good agreement between
the two plots. The small difference may be attributed to slightly
different parameter values (e.g. axial mass) or to the
approximation assumed in the Marteau model where the terms of
order ${|\vec p|\over M}$ ($\vec p$ is target nucleon momentum)
are omitted from the hadronic tensor. In the detailed comparison
of
 several MC codes predictions which was presented by Zeller during her
NuInt02 talk \cite{zeller} even bigger differences were shown.\\

The results for the scattering on the oxygen nucleus are shown in
Fig.~2.  The NUX+FLUKA events are classified
 based on the particles leaving the primary vertex and as a
result the FSI is not taken into account. To be more specific for
the plot of the CC quasi-elastic scattering cross section only
events with proton and muon produced in the primary vertex are
considered. On the other hand the plot for the Marteau model was
obtained without the RPA corrections. In other words two free
Fermi gas models are compared.


The difference between the plots seen in Fig.~2 has at least two
reasons. Firstly it is inherited from the difference seen in
Fig.~1. Secondly it can be caused by different distributions of
the target nucleons momenta assumed in the two models. In the
Marteau model a quadratic distribution with sharp cut at $k_F=225
MeV$ is assumed while in the NUX+FLUKA event generator
a smooth distribution is used \cite{salla}.\\

To be able to compare theoretical predictions with experiment such
nuclear  effects as the reinteraction or the $\Delta\to NN$
channel must be taken into account.

The plots in  Fig.~3 were obtained by using the particles seen in
the final state to select the CC quasi-elastic scattering events.
This means that in the NUX+FLUKA generated event the primary
vertex might be classified as either quasi-elastic or single pion
production but not only. However, we neglected rather small
contributions from other primary vertex kinds because they are
not included in the Marteau model. The plot obtained for the
Marteau-Model includes RPA corrections.

The comparison of plots found in figures 2 and 3 leads to the
conclusion that nuclear effects increase the cross section.

This increase comes from an extra contribution from pion
absorption in the case of the NUX+FLUKA generator and from the
non-pion decays of the $\Delta$ resonance in the case of the
Marteau model. It is interesting that two contributions of so
different origin are of the same order of magnitude.

The difference between these two corrections rises with the
neutrino energy reaching about $30\%$ at 2GeV and causes the
cross-section produced by the NUX+FLUKA generator to be slightly
higher than the one obtained in the Marteau model.

\section{SINGLE PION PRODUCTION}

\rys{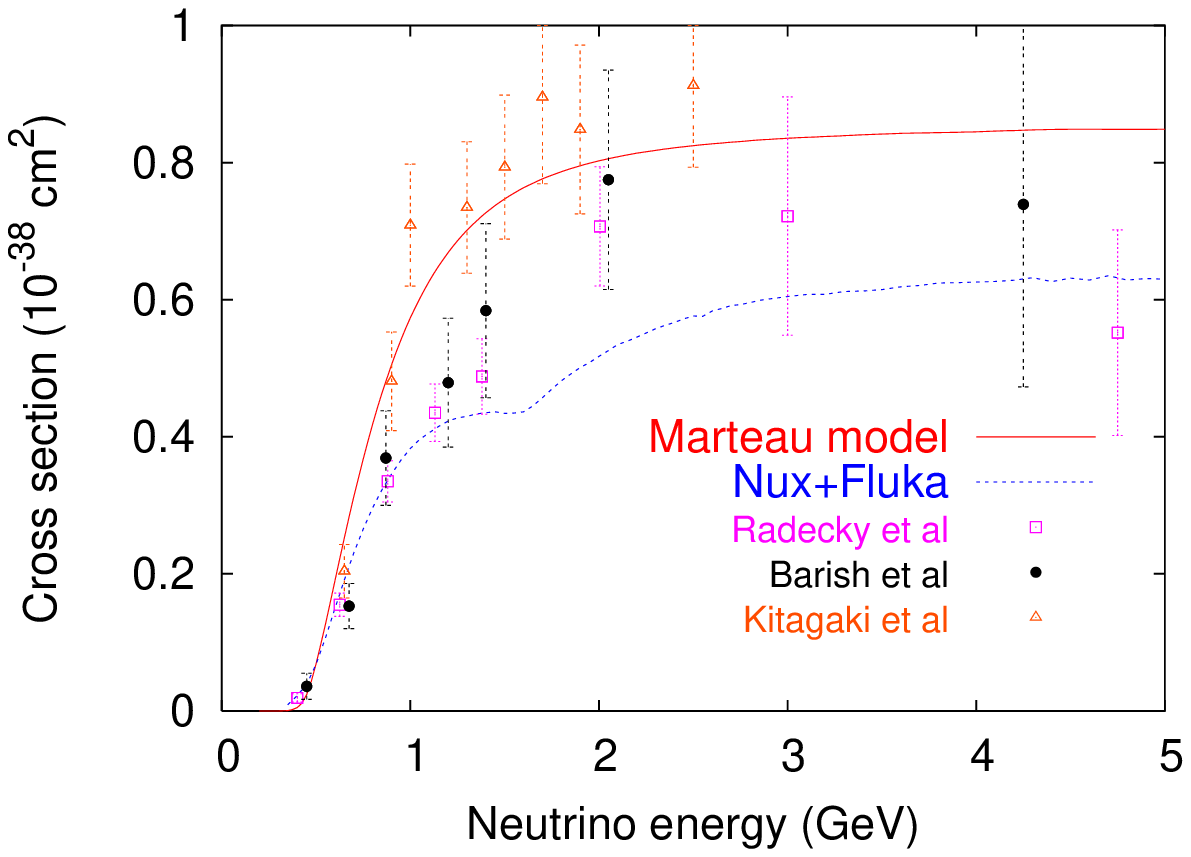}{Total cross section for single $\pi ^+$
production on free protons. Experimental points are taken from
\cite{pion_exp},\cite{kitagaki}. \label{com_pion_free}}

\rys{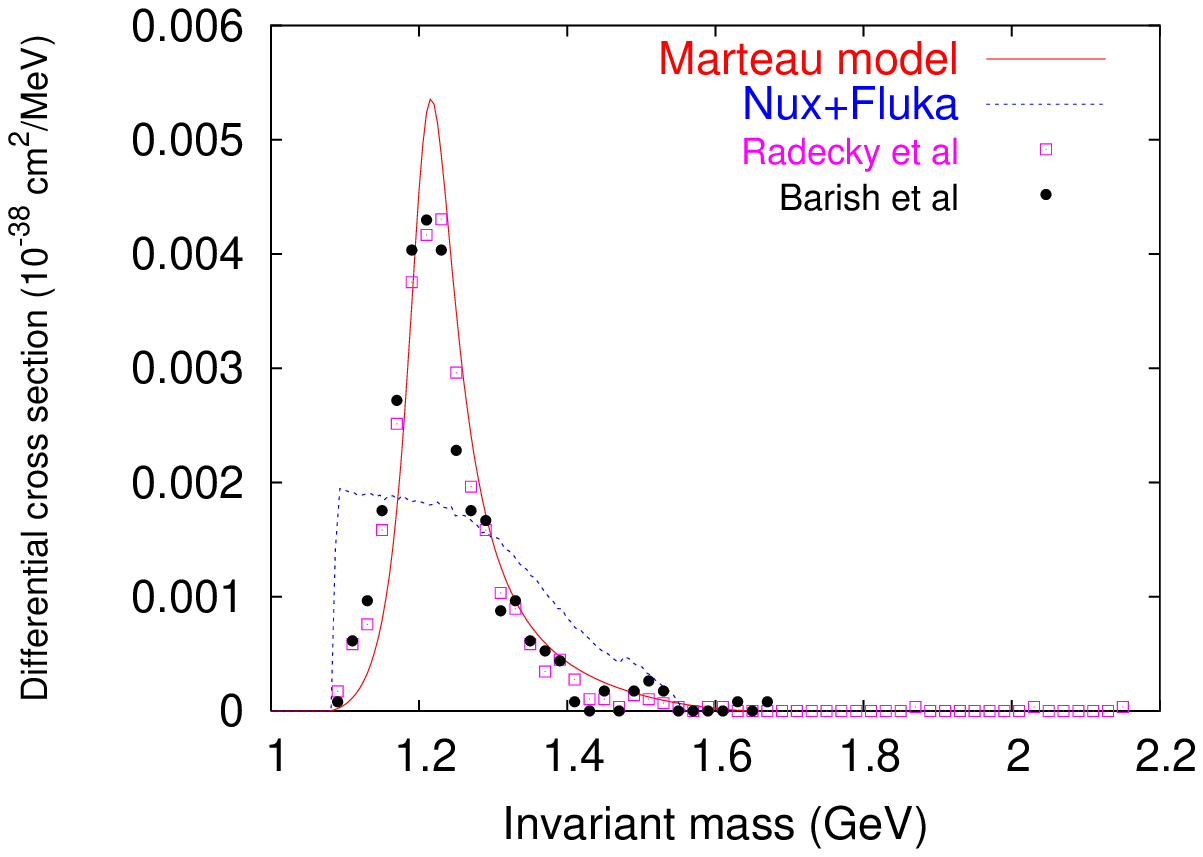}{Normalized invariant hadronic mass
distribution in single $\pi^+$ production on free protons. See
explanation in the text.\label{com_pion_mass_free}}

\rys{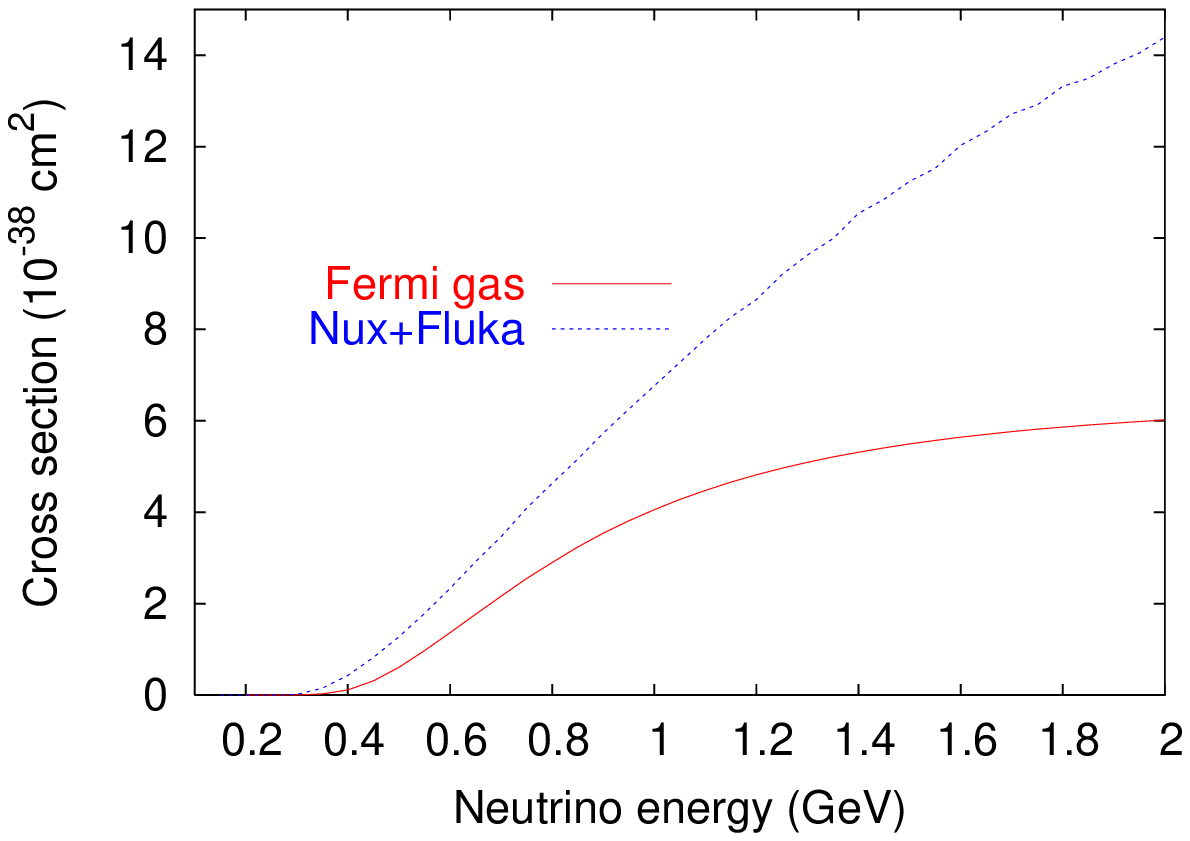}{Total cross section for single $\pi$
production on $^{16}O$ before Final State Interactions.
\label{com_pion_FG}}

\rys{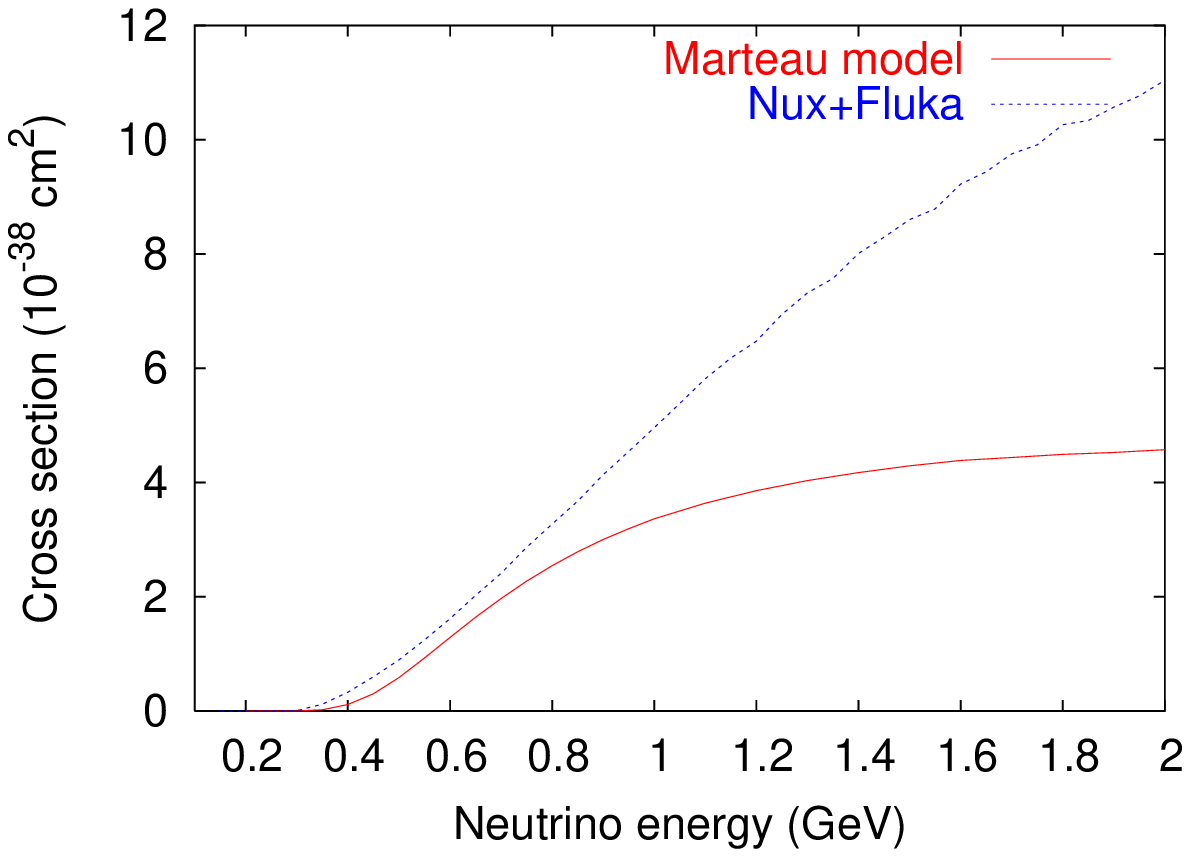}{Total cross section for single $\pi$
production on $^{16}O$ (Final State Interactions
included).\label{com_pion_RPA}}

\rys{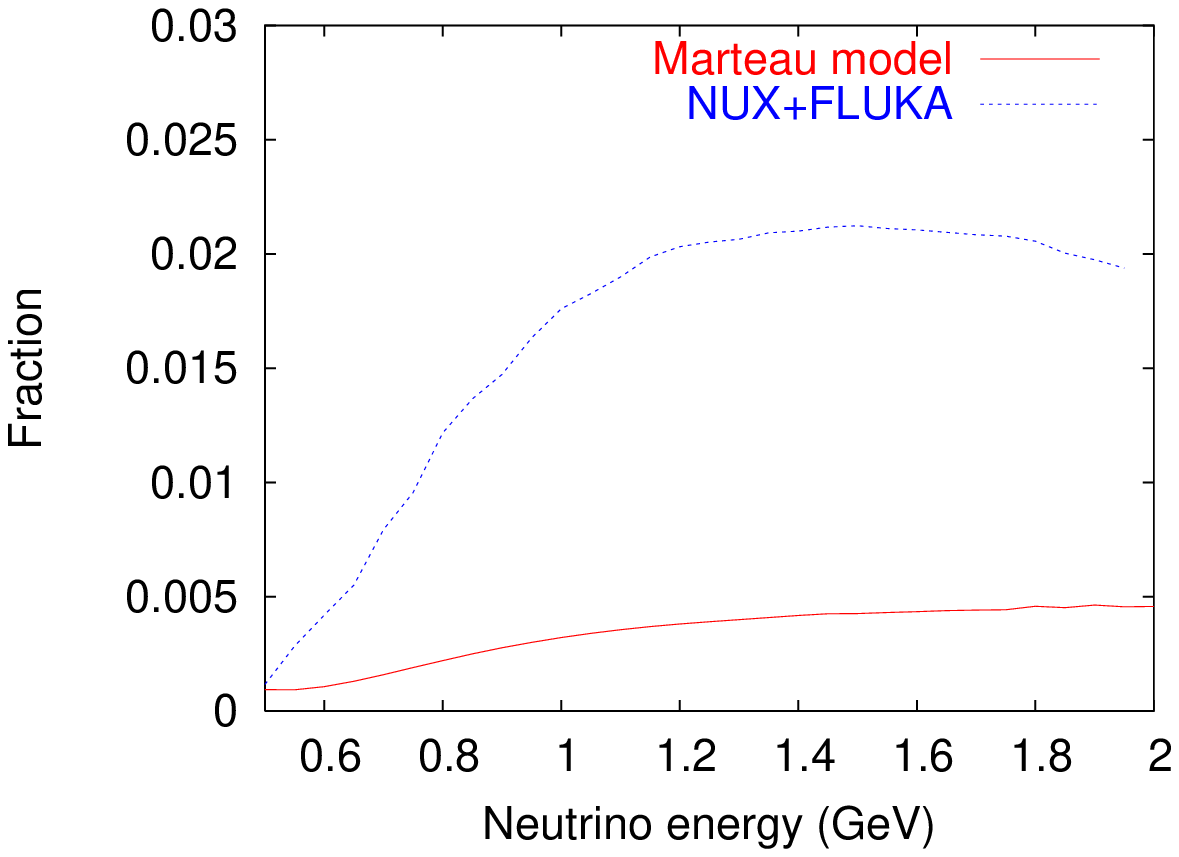}{Fraction of pions produced from quasielastic
primary vertex due to nuclear effects.}

Predictions for single $\pi$ production on free nucleons are
compared only for the $\nu p\rightarrow \mu^- p\pi^+$ channel. In
this channel most of pions are produced through $\Delta^{++}$
excitation. In the experimentally measured hadronic invariant mass
distribution a resonance peak is clearly seen \cite{pion_exp}.  In
the case of $\nu n\rightarrow \mu^- n\pi^+$ and $\nu n\rightarrow
\mu^- p\pi^0$ channels the resonance peak is smeared out which
indicates that the dynamics is
more complicated.\\

In Fig.~4 we compare the total cross sections predicted by both
models with the experimental points. While both curves agree
roughly with the results of Barish\cite{pion_exp}, the data
points from \cite{kitagaki} seem to favour the Marteau model.

In Fig.~5 a similar comparison is made for the normalized
hadronic invariant mass distributions. The neutrino energy has
been fixed at 1~GeV in both models because the experimental points
were taken from \cite{pion_exp} where most of the events came
from neutrinos of energy around 1GeV.
 In order to compare only distribution's shapes
all the plots have been normalized to yield the same total cross
section value (that of the Marteau model at neutrino energy
$1GeV$). The shape obtained in the Marteau model agrees well with
the data points which further justifies the hypothesis that most
of the pions come from the $\Delta^{++}$ decay.  The disagreement
between the shape obtained from NUX+FLUKA event generator and the
data points is not surprising in this context as it is known not
to contain any explicit resonance contribution.

In the next two figures the total cross sections of the pion
production on oxygen are presented. Contributions from all three
pion production channels were added here because in the Marteau
model they are related to each other by the isospin Clebsch-Gordan
rules and they cannot be tested independently. In Fig.~6 we
classified the events based on the particles leaving the primary
vertex. There is a major disagreement between the two curves:
NUX+FLUKA predicts cross section to be much higher then the one
obtained from the Marteau model.

At neutrino energy $2 GeV$ the difference is already by a factor
of 2 and it is seen that in the Marteau model the cross section
saturates at much lower energy. The two main reasons for this
discrepancy are the following:
\medskip

1) in the Marteau model pions are produced only via the $\Delta$
excitation with other (non-resonant, higher resonances)
contributions neglected. Therefore the cross sections are
underestimated
\smallskip

2) as shown in \cite{zeller} the NUX+FLUKA predictions in $\nu
n\rightarrow \mu^- n\pi^+$ and $\nu n\rightarrow \mu^- p\pi^0$
channels are significantly overestimated.\\

In Fig.~7 the cross sections for single pion appearance in the
final state are shown. Prediction of the Marteau model is again
much lower. Analysis of \cite{zeller} shows that already at
neutrino energy $2 GeV$ NUX+FLUKA predict cross section higher
then saturation plateau of approaches based on Rein-Sehgal model.
By adding three contributions one can expect that the plateau
should be at approximate value $10\cdot 10^{-38} cm^2$.\\

Finally we considered the fraction of pions that are produced due
to the Final State Interactions. In Fig.~8 we show the predictions
of both approaches. The value of this fraction obtained in the
NUX+FLUKA scheme is four times larger then the one calculated in
the Marteau model.

This big difference can be discouraging but it is instructive
that questions like this can be posed and answered within both
numerical and analytical frameworks.

 We hope that the discussion about numerical and analytical
 approaches to the description of nuclear
effects in neutrino interactions can lead to improvements in
existing MC codes.

\end{document}